\documentclass[%
 reprint,
 amsmath,amssymb,
 aps,prl
]{revtex4-2}

\usepackage{graphicx}
\usepackage{dcolumn}
\usepackage{bm}
\usepackage[dvipsnames]{xcolor}
\usepackage{soul}

\newcommand{\comment}[1]{} 

\begin{document}

\preprint{APS/123-QED}

\title{Competition on the edge of an expanding population}

\author{Daniel W. Swartz}
\author{Hyunseok Lee}
\author{Mehran Kardar}
\affiliation{Department of Physics, Massachusetts Institute of Technology, Cambridge, Massachusetts 02139, USA}
\author{Kirill S. Korolev}
\affiliation{Department of Physics, Graduate Program in Bioinformatics and Biological Design Center, Boston University, Boston, Massachusetts 02215, USA}

\date{\today}

\begin{abstract}
    
    In growing populations, the fate of mutations depends on their competitive ability against the ancestor and their ability to colonize new territory. Here we present a theory that integrates \textit{both aspects} of mutant fitness by coupling the classic description of one-dimensional competition~(Fisher equation) to the minimal model of front shape~(KPZ equation). We solved these equations and found three regimes, which are controlled solely by the expansion rates, solely by the competitive abilities, or by both. Collectively, our results provide a simple framework to study spatial competition. 
\end{abstract}

\maketitle

Propagating fronts are a ubiquitous feature of spatially extended systems. Examples include the spread of an invasive species in an ecosystem~\cite{murray2002mathematical}, the spread of a ferromagnetic phase across a magnet~\cite{eisler2016universal}, the spread of a high fitness allele through a population~\cite{fisher1937wave}, or even the propagation of a flame front ~\cite{zeldovich1959theorv}.
These and other applications have stimulated a sustained effort to construct and analyze coarse-grained models of traveling reaction-diffusion waves~\cite{murray2002mathematical,panja2004effects,tang1980propagating,kolmogorov1937moscow,van2003front}. By now, we have a good general understanding of one-dimensional waves, but two and higher dimensions pose numerous challenges because of the interplay between the dynamics along the wave front and the shape of the wave front itself.

Growing microbial colonies provide an excellent experimental system to study the two-way coupling between the shape of the colony edge and the spatial distribution of different genotypes in the population~\cite{hallatschek2007genetic,hallatschek2008gene,korolev2012selective}. At the same time, microbial colonies also serve as useful model systems for tumor growth and geographic expansions of plants and animals~\cite{gidoin2019range}. Hence, there has been a lot of interest in understanding the spatial competition between two different genotypes, say a mutant and an ancestor at the edge of a growing colony~\cite{hallatschek2010life,excoffier2009genetic}. 


Theory of competition at the edge of colony is not well established yet. 
Most computational studies rely on numerical simulations of complex microscopic models, and therefore can draw few general conclusions about possible outcomes of spatial competition~\cite{nadell_emergence_2010,ghaffarizadeh2018physicell,metzcar2019review}. 
Focusing on relevant symmetries, a recent theoretical approach~\cite{horowitz2019bacterial} provides a framework for morphologies in competitive growth.
A corresponding geometric interpretation also classifies the small number of distinct colony growth patterns and competitive outcomes~\cite{lee2022slow,korolev2012selective,hallatschek2010life}. The theoretical models, however, are agnostic to the mechanism of competition and assume the knowledge of  emergent properties, such as the invasion velocity of the mutant. In consequence, their utility is rather limited because they cannot, for example, predict the winner of the competition given the microscopic qualities of the mutant and the ancestor. 

At the core of our approach is the assumption that the state of the colony is well-described by two quantities: the spatial extent or ``height'' of the colony~$h(x,t)$ and mutant fraction~$f(x,t)$, which change along the colony front~($x$-coordinate) and with time~$t$~\cite{horowitz2019bacterial}. For simplicity, we consider only planar fronts, where~$h$ is simply the distance by which the colony expanded from the inoculation site. Thus, we treat the colony edge as a thin interface. This is a reasonable approximation because the growth region extends only a few cell widths into the colony and any successful mutant has to emerge near the colony edge; otherwise it is crowded out of the growth zone and remains trapped in the colony bulk.

The dynamical equations for~$h(x,t)$, and~$f(x,t)$ emerge naturally as generalization the well-studied limits of the Fisher-Kolmogorov-Petrovsky-Piskunov~(FKPP) equation~\cite{fisher1937wave,kolmogorov1937moscow} for one dimensional competition (no variation in~$h$); and the Kardar-Parisi-Zhang~(KPZ) equation~\cite{kardar1986dynamic} for interface growth (with no variation in~$f$).

The shape of the colony front is described by the generalized KPZ equation~\cite{kardar1986dynamic}:
\begin{equation}
    \frac{\partial h}{\partial t} = v_0 + \frac{v_0}{2}\Big(\frac{\partial h}{\partial x}\Big)^2 + D_h \frac{\partial^2 h}{\partial x^2} + \alpha f,
    \label{eqn:h}
\end{equation}
\noindent where the first two terms express the \textit{isotropic expansion} of the colony with velocity~$v_0$ along the local normal to the front; the third term encodes curvature relaxation, and the fourth term accounts for the difference in the expansion velocities of the mutant and the ancestor~\cite{horowitz2019bacterial}. We used a linear interpolation between the two velocities because most of our result are obtained ``to the first order'' in the differences between the two competitors. Higher order terms (such as $\alpha f(\partial h/\partial x)^2$) are similarly ignored. 

The dynamics of the mutant fraction~$f$ is described a modified FKPP equation~\cite{fisher1937wave,kolmogorov1937moscow}:
\begin{equation}
    \frac{\partial f}{\partial t} = s(f) f (1-f) + D_f \frac{\partial^2 f}{\partial x^2} + v_0 \frac{\partial h}{\partial x}\frac{\partial f}{\partial x},
    \label{eqn:f}
\end{equation}
\noindent where the first term accounts for differences in local reproduction rates, the second term accounts for spatial rearrangements due to motility or population fluxes generated by the expansion dynamics, and the third term is our addition to describe ``passive'' changes in~$f$ due to the motion of a tilted interface~\cite{george_chirality_2018,horowitz2019bacterial,chu2019evolution}. This last term can be easily understood by considering what happens to a cell when a tilted interface moves in the direction of its normal: such a cell moves both ``vertically,'' i.e. in the~$h$-direction and ``horizontally,'' i.e. along the front. 

Without the coupling to~$h$, the FKPP equation is the classic model of spatial competition between two species or two genotypes in one dimension. Its asymptotic solutions are known as traveling waves because they have the form~$f(x,t) = f(x-ut)$, where~$u$ is the invasion velocity of the mutant. 
In the following, we determine how the coupling between~$h$ and~$f$ affects~$u$ by solving Eqs.~\eqref{eqn:h} and \eqref{eqn:f} numerically using MATLAB's \texttt{pdepe}. We then develop an analytical theory that not only quantitatively matches the simulations, but also provides deep insights into the existence of three distinct regimes of spatial competition.

The solutions of the FKPP equation are broadly classified into so-called ``pulled'' and ``pushed'' waves depending on how the selective advantage~$s$ depends on mutant frequency~$f$. Pulled waves are dominated by the dynamics at leading edge, and the invasion velocity can be obtained by linearizing the FKPP equation for small~$f$; the resulting `Fisher velocity' is given by~$u_{\mathrm{\textsc{f}}}=2\sqrt{D_fs(0)}$ (see Refs.~\cite{van2003front,fisher1937wave,kolmogorov1937moscow,murray2002mathematical}). In contrast, the velocity of pushed waves depends on the values of~$s$ at all~$f$, and cannot be, in general, computed analytically except for some exactly solvable models such as with~$s(f) = s_0(f-f_0)$~\cite{fife1977approach,roques2012allee}. For this model, pushed waves occur for~$f_0\in(-0.5,0.5)$ with~$u=\sqrt{D_f s_0/2}(1-2f_0)$; the waves are pulled for~$f_0<-0.5$. Behavior for~$f_0>0.5$ is analyzed by changing variables from~$f$ to~$1-f$.

\begin{figure}
    \centering
    \includegraphics[width=0.5\textwidth,height=7cm]{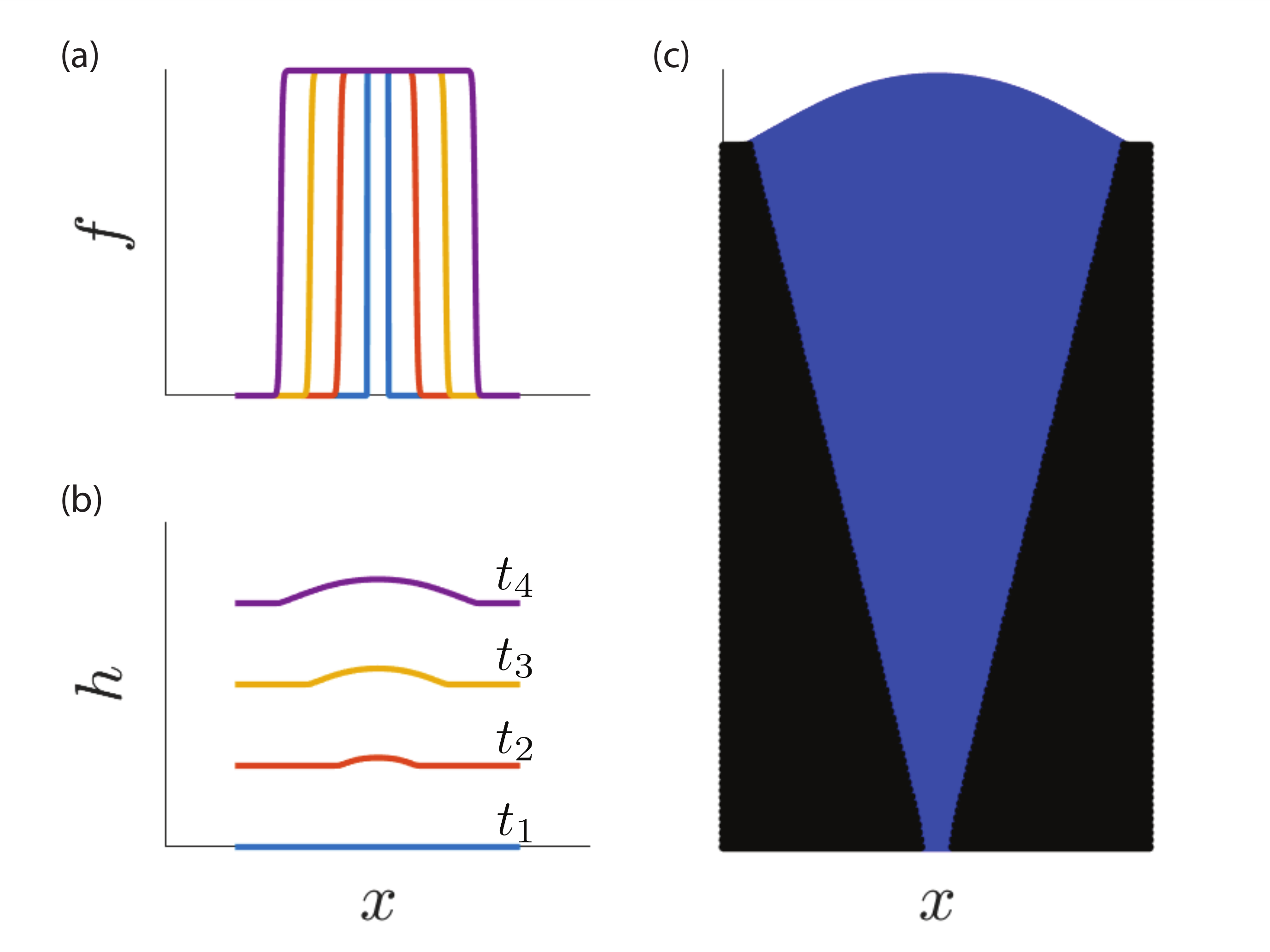}
    \caption{Sample simulation results generated by solving Eqs.~\eqref{eqn:h} and \eqref{eqn:f} numerically  in the pulled wave  regime $s(f) = s_0$. {\bf (a)} Profiles of mutant frequency  $f(x,t)$ at five equally-spaced time slices labeled by color. {\bf (b)} Height profiles $h(x,t)$ taken at the same five time points as in panel (a) with the same color convention. Starting from a flat initial condition, the height field develops a nontrivial morphology through a dependence of the growth rate of $h$ on the mutant frequency $f$. {\bf (c)} The spatial distribution of the two competitors is visualized by plotting successive solutions of the height field $h$ and coloring each point according to the value of $f$ at the corresponding~$x$ and~$t$.}
    \label{fig:modelsketch}
\end{figure}

Without loss of generality, we assume that the mutant has a local fitness advantage~$s(f)>0$, and first consider pulled waves with~$s(f)=s_0>0$. The emergence of a traveling wave of $f(x,t)$ is apparent in Fig.~\ref{fig:modelsketch}(a). The corresponding~$h(x,t)$, however, is not a simple traveling wave~(Fig.~\ref{fig:modelsketch}(b)). The colony front is composed of a curved portion dominated by the `mutant' and a flat front dominated by the `ancestor'. While the transition point between these two regimes advances with the same velocity~$u$, the overall shape of the curved portion depends on both time and the co-moving coordinate~($x-ut$) because the growth dynamics encoded by~$\alpha$ persist even after the mutant displaces the ancestor.

To understand how the invasion velocity~$u$ is affected by the coupling to height, we computed it numerically at different values of~$\alpha$. For~$\alpha=0$, the equations are effectively decoupled because genetic variation along the front does not create any disturbances in the front shape, which remains~$h=v_0t$ for all~$x$. Mutant is faster than the ancestor when~$\alpha>0$ and slower otherwise. While the latter case may seem paradoxical, it has actually been observed experimentally~\cite{lee2022slow}.

\begin{figure}
    \centering
    \includegraphics[width = 0.45\textwidth]{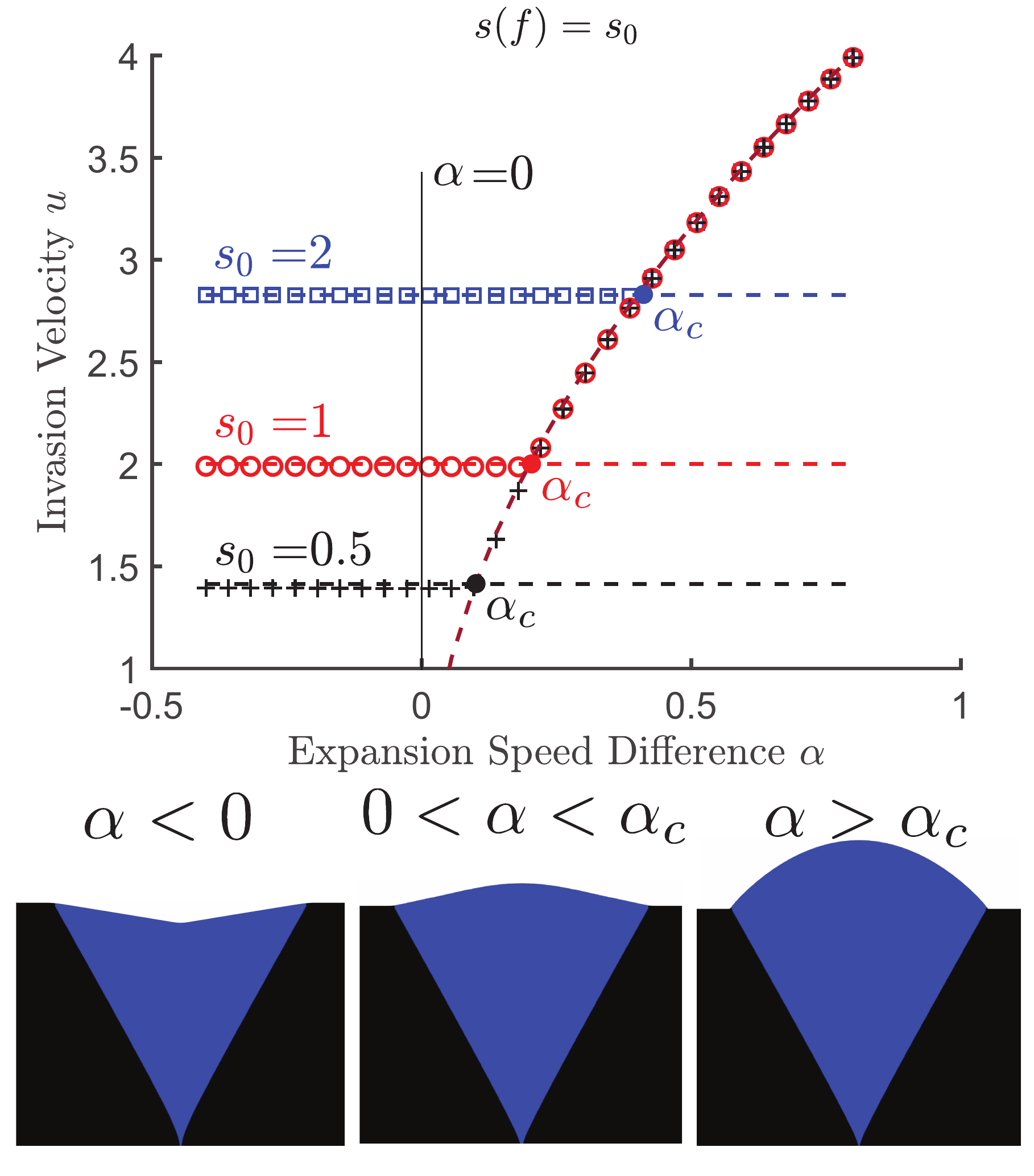}
    \caption{Invasion dynamics in pulled waves. (Top): Invasion velocity shows two regimes with dependence on either~$\alpha$ or~$s$.  The horizontal dashed lines are the predicted `Fisher velocities'. The curved dashed line is~$u=\sqrt{2\alpha v_0}$. For each value of $s_0$ the filled in circle shows the location of the transition point $\alpha_c$ between the composite and circular arc morphologies. (Bottom): Depending on~$alpha$, there are three distinct colony morphologies. When $\alpha<0$ the front shape is a V-shaped dent. When $0<\alpha<\alpha_c$ the morphology is a composite bulge consisting of a central circular arc transitioning to a constant slope at the bulge edges. When $\alpha>\alpha_c$ the front is entirely a circular arc. (Parameters used are $v_0 = 10, D_f = D_h = 1$.)}
    \label{fig:pulled_alpha_sweep}
\end{figure}

Simulation results are shown in Fig.~\ref{fig:pulled_alpha_sweep}, with different markers denoting  different values of $s_0$. We immediately observe that the data falls into two regimes: For small~$\alpha$, the invasion velocity is a constant, which depends on~$s_0$, but not~$\alpha$. For large~$\alpha$, the situation is reversed: the velocity depends on~$\alpha$, but not on~$s_0$. Thus, there appears to be two distinct regimes: one mediated by local competition described by the FKKP equation, and one mediated by the expansion rates in the KPZ equation.

We tested this hypothesis by comparing solutions of the uncoupled FKKP and KPZ equations to the results  in Fig.~\ref{fig:pulled_alpha_sweep}. For small~$\alpha$, there is perfect agreement between the observed values of~$u$ and the expected Fisher velocity~$u_{\mathrm{\textsc{f}}}=2\sqrt{D_fs_0}$.In hindsight, this may not be too surprising since pulled waves are controlled by the dynamics at the leading edge, where~$h$ is flat and the coupling term is the FKPP equation vanishes. However, the influence of growth velocity differences manifests dramatically in the shape of the front, which changes from a V-shaped dent at negative~$\alpha$ to a composite bulge for positive~$\alpha$; see Fig.~\ref{fig:pulled_alpha_sweep}.

For large~$\alpha$, the simulations match
\begin{equation}
    u =\sqrt{2\alpha (v_0+\alpha)} \approx \sqrt{2\alpha v_0},
    \label{eqn:circarcspeed}
\end{equation}
\noindent which can be obtained from the KPZ equation by analogy with the equal-time argument in Ref.~\cite{korolev2012selective} and the geometric theory in Ref.~\cite{lee2022slow}. In this regime, the mutant forms a circular bulge~\footnote{Since the KPZ equation approximates isotropic growth only to first order, the height field actually has a parabolic shape:~$h(x,t) = (v_0 + \alpha)t - \frac{x^2}{2(v_0+\alpha)t}$.} of radius~$(v_0+\alpha)t$, while the ancestor has a flat front at height~$v_0t$. 
These two curves intersect at point whose $x$-coordinate moves with velocity~$u=\sqrt{2\alpha(v_0+\alpha)}\approx\sqrt{2\alpha v_0}$.
The transition between the two regimes occurs at a critical value of~$\alpha_c=2s_0 D_f / v_0$ when the velocity of the circular bulge exceeds the Fisher velocity. This agrees with the general observation that a faster moving solution typically controls the behavior of a traveling wave~\cite{van2003front}. 

\begin{figure}
    \centering
    \includegraphics[width=0.5\textwidth]{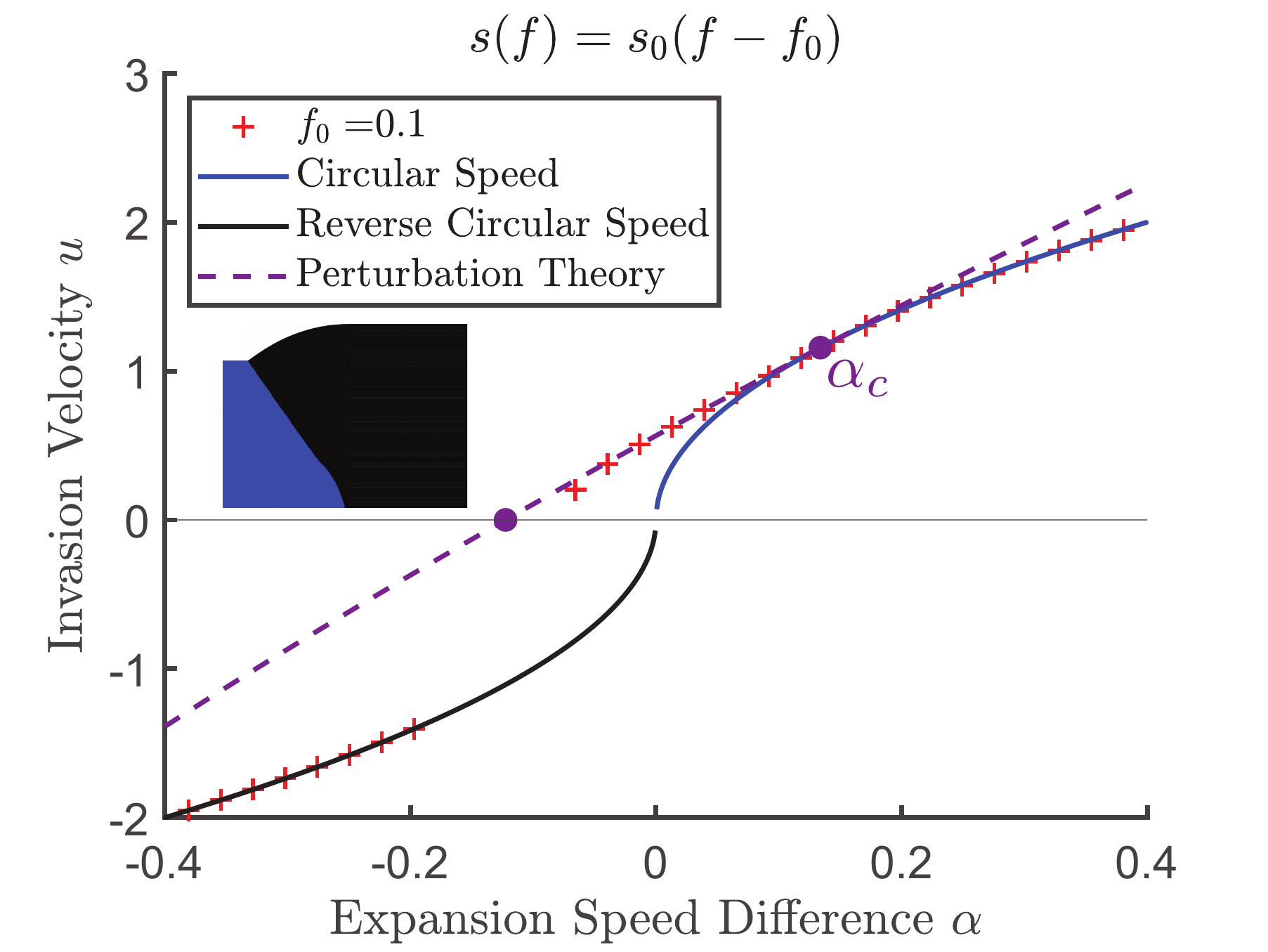}
    \caption{Invasion dynamics in pushed waves.  Simulation results are shown with symbols. When $\alpha$ is sufficiently negative, the invasion velocity is negative~($u<0$), i.e. the mutant is invaded by the wildtype. In this regime, the data are well-described by a circular arc velocity $-\sqrt{-2\alpha v_0}$. For large and positive~$\alpha$, we again see the a circular arc velocity, $u=\sqrt{2\alpha v_0}$. For values of $\alpha$ near 0, the perturbative treatment described in the text results in the nontrivial dependence on $\alpha$ depicted by the dashed line. The two large dots show where our theory predicts a transition in the behavior of the invasion velocity/morphology. The dot labeled $\alpha_c$ marks the transition between the composite bulge and circular arc morphology, as discussed in the pulled case. The leftmost dot shows where the predicted velocity from perturbation theory would vanish, which we interpret as the reversal of the selective pressure with the mutant  now invaded by the wildtype. Inset in the figure is a sample morphology which arises when the mutant is invaded by the wildtype~($u<0$). The simulation shown is initialized as a half-space where the left half is occupied by mutant and the right by wildtype. Parameters are $v_0 = 5, s_0 = 1, f_0 = 0.1,D_f=D_h=1$. We were unable to carry out simulations through the actual transition point because the timescale on which the wave attained its steady-state velocity grows dramatically with $\alpha - \alpha_c$, reminiscent of the phenomenon of critical slowing down. Simulations suggest that when $\alpha$ is below this negative transition point, there is a sudden jump in the invasion velocity. 
}
    \label{fig:pushed_alpha_sweep}
\end{figure}

The above results for pulled waves are surprising from both mathematical and  biological perspectives. Mathematically, it is surprising that the invasion velocity~$u$ is controlled by only one of the equations, i.e. there is no two-way coupling. Biologically, it seems counter-intuitive that, no amount of disadvantage in the expansion velocity~($\alpha<0$) can overcome the competitive advantage~($s_0$). To see whether these conclusion hold more generally, we carried out equivalent simulations for pushed waves with~$s(f) = s_0(f-f_0)$. The results are shown in Fig.~\ref{fig:pushed_alpha_sweep}.

Compared to pulled waves, there are two major differences. 
First,~$u$ changes sign when~$\alpha$ becomes sufficiently negative. 
In this case, the ancestor invades a more competitive mutant~($s_0>0$) because it has a much large expansion velocity. 
The invasion proceeds with a circular bulge of the ancestor advancing at velocity~$u=\sqrt{2|\alpha|v_0}$. 
Second, there is no regime where~$u$ does not depend on~$\alpha$. 
Thus, pushed waves exhibit two regimes: one dominated by the height profile, which occurs for both positive and negative values of~$\alpha$, and the other with a two-way interplay between the KPZ and the FKPP dynamics at intermediate~$\alpha$. 
The former regime is identical to the one discussed in the context of pulled waves, so we proceed to analyze the latter using a perturbative approach. 

Unlike in pulled waves, the variations throughout the front are as important as dynamics at its leading edge, such that the couplings cannot be  neglected. Instead, we can resort to a perturbative treatment of the coupling term
$v_0 h' f'$ (with $h'\equiv \partial h/\partial x$
and $f'\equiv \partial f/\partial x$) in Eq.~\eqref{eqn:f}, using an approach detailed in Refs.~\cite{meerson2011velocity,paquette1994structural,mikhailov1983stochastic,rocco2001diffusion}. 
The coupling term~$v_0f'h'$ is non-zero only in the interval when~$f$ changes from~$0$ to~$1$. In this region,~$h'$ changes from~$0$ to a maximum slope which we denote as~$\sigma$. 
The scale of the perturbation is thus set by
$v_0\sigma$.

The perturbative scheme proceeds as follows:
In the absence of coupling, the  profile for $f(z=x-ut)$ is $f^{(0)}(z) = {1}/({1+e^{z/a}})$,
with $a = \sqrt{{2 D_f}/{s_0}}$ and $u_0 = \sqrt{{s_0 D_f}/{2}}(1-2f_0)$. 
Substituting $f^{(0)}(z)$ into Eq.~\eqref{eqn:h} 
leads to a non-linear differential equation whose solution provides the first order profile $h^{(1)}(z=x-ut)$.
As described in the SI the non-linear equation can be solved exactly via a Cole-Hopf transformation, resulting in a complicated form for $h^{(1)}(z=x)$ that depends on $v_0$, $D_h$ and $\alpha$. Qualitatively, this height profile is a sigmoidal curve that changes from $v_0t$ in the region $f\to0$ ($h'\to0$), to $v_0t+\sigma z$ as $f\to1$ ($h'\to\sigma$), with the limiting slope given by
\begin{equation}
    -u \sigma = \alpha + {v_0}\sigma^2/2\,.
    \label{eqn:sigma}
\end{equation}

After substituting $h'^{(1)}(z)$ into Eq.\eqref{eqn:f}, the methodology described in
Refs.~\cite{meerson2011velocity,paquette1994structural,mikhailov1983stochastic,rocco2001diffusion}
can be used to compute the first order 
correction to the invasion velocity, leading to the perturbative form 
\begin{equation}
    u = u_0 - \kappa v_0 \sigma + O(v_0^2\sigma^2)\,,
    \label{eqn:u}
\end{equation}
\noindent where~$u_0$ is the unperturbed velocity for~$\alpha=0$.

To coefficient $\kappa$ in Eq.~\eqref{eqn:u} is obtained numerically as ratio of integrals that depend on the function $h^{(1)}(z)$ (see SI). In general, the solution is complex, but it can be simplified in two limiting cases.

\begin{figure}
    \centering
    \includegraphics[width=0.45\textwidth]{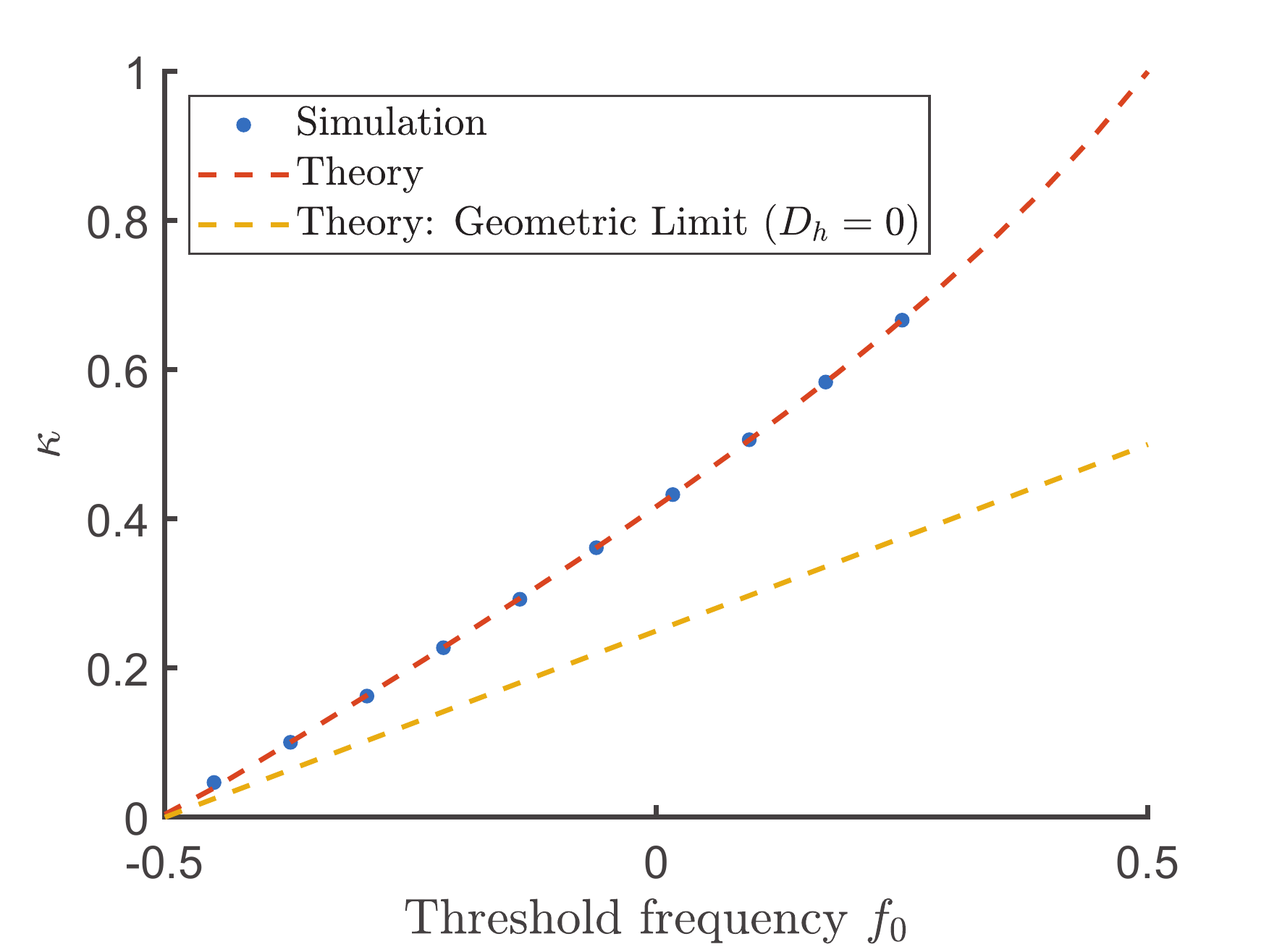}
    \caption{Numerical results showing the dependence of the coefficient $\kappa$ as defined in Eq.~\eqref{eqn:u} on $f_0$. The numerical values of $\kappa$ were obtained by fitting measured invasion velocities as functions of $\alpha$ in the limit $\alpha \to 0$. The best-fit slope is then used to obtain $\kappa$ in Eq.~\eqref{eqn:u}. The red dashed line is the theoretical prediction of our perturbaive analysis for the value of $D_h$ used in simulation at small $\alpha$. The yellow dashed line is the theoretical value of $\kappa$ when $D_h =0$. Parameters are $v_0=10,D_h=D_f=1,s_0=2$.}
    \label{fig:kappavsf0}
\end{figure}

By setting~$D_h=0$, equation for $h^{(1)}(z=x)$ becomes first order, and its solution simplifies the evaluation of all downstream integrals. This limit corresponds to the geometric description in which the profile simply advances along the local normal without further relaxation and yields $\kappa_\text{geom.} = \frac{1}{4}(1+2f_0)$. From Fig.~\ref{fig:kappavsf0}, it is clear that the geometric limit is not very accurate for our simulations with~$D_h=1$. Nevertheless, it captures the qualitative changes in~$\kappa$ including the transition to pulled waves at~$f_0=-0.5$ at which~$\kappa$ must vanish. Thus, our theory recapitulates the finding that the invasion speed becomes insensitive to the morphology when the wave becomes pulled. 

Another useful approximation is obtained by neglecting the nonlinear term in Eq.~(\ref{eqn:h}), which is justified for small~$\alpha$ because~$\partial h/\partial x \propto \alpha$. In this case, we have to evaluate the downstream integrals numerically, but obtain a perfect agreement with the simulations at least when~$\alpha$ is small; see Fig.~\ref{fig:kappavsf0}.

Since perturbation theory can be applied to both positive and negative~$\alpha$, we can determine the location of the transition from a right-moving wave~(mutant taking over) to the left-moving wave~(ancestor taking over), which occurs at~$u=0$, i.e for~$\alpha=-{u_0^2}/({2v_0\kappa^2})$. Beyond this point, the perturbation theory is no longer applicable, and simulations show that~$u$ jumps discontinuously from~$u=0$ to the value corresponding to a left-moving circular arc as discussed above.

Microbes, cancer cells, and invasive species often spread across space forming a continuous two-dimensional populations. Nearly all successful mutations occur near the frontier of growth where mutants have a high chance to colonize new territory. Understanding this establishment process is complicated as competition occurs at a moving interface whose shape is also constantly changing. Here, we couple a model of surface growth, namely the KPZ equation, to a model of competition, namely the generalized FKPP equation. The combined model faithfully describes recent observations of nontrivial colony morphologies near emerging mutants. Moreover, it elucidates how colonization rate and local competitive strength affects the fate of the mutation. We find three distinct scenarios of the takeover by a mutant: \textbf{(1)} Mutants can establish purely due to  higher colonization rate producing a characteristic circular bulge whose shape depends only on the ratio of the colonization rates of the mutant and the ancestor. \textbf{(2)} The mutant can also be fixed purely due to its higher competitive ability. Such mutants form either a dent or a non-circular bulge depending on whether their colonization rate is lower or higher than that of the ancestor. \textbf{(3)} There is also a scenario in which both the colonization rate and the competitive ability play a role.

These distinct types of mutant takeover are controlled by competition dynamics encoded in the FKPP equation. Surprisingly, the outcomes described above depend on whether the FKPP equation admits pulled traveling waves driven by the growth dynamics at low mutant densities or by pushed traveling waves driven by the growth dynamics throughout the mutant-ancestor interface. These conclusions are supported by numerical simulations and analytical perturbation theory. 
Taken together our results not only elucidate many subtleties associated with mutant establishment, but also pave the way for a more parsimonious and universal description of evolutionary and ecological processes in growing populations that is also very amenable to theoretical analyses. 

K.S.K was supported by the NIGMS grant 1R01GM138530-01.
M.K. acknowledges support from NSF through grant DMR-1708280.
D.W.S. is supported by the MathWorks School of Science fellowship.
H.L. is supported by Sloan Foundation through grant G-2021-16758.

\bibliography{bibliography}{}

\end{document}